\documentclass[preprintnumbers,
amsmath,amssymb,floatfix,10pt,prd,onecolumn,
superscriptaddress,longbibliography,nofootinbib]{revtex4-2}
\usepackage{bm}
\usepackage{amsfonts}
\usepackage{latexsym}
\usepackage{amsmath}
\usepackage{textcomp}
\usepackage{float}
\usepackage{booktabs}
\usepackage{dcolumn}
\usepackage{dsfont}
\usepackage{ragged2e}
\usepackage{epsfig}
\usepackage[dvipsnames]{xcolor}
\usepackage{hyperref}
\hypersetup{           
	colorlinks=true,                
	breaklinks=true,                
	urlcolor= OrangeRed,                
	linkcolor= magenta,                
	bookmarksopen=false,
	filecolor=black,
	citecolor=red,
	linkbordercolor=blue}
\usepackage{graphicx}
\usepackage{orcidlink}
\usepackage[T1]{fontenc}

\begin{document}
	\title{New Agegraphic Dark Energy Driven Reconstruction of \boldmath{$f(Q)$} Gravity and its Cosmological Implications}
	
   \author{Rajdeep Mazumdar \orcidlink{0009-0003-7732-875X}}
	\email{rajdeepmazumdar377@gmail.com}
	\affiliation{%
		Department of Physics, Dibrugarh University, Dibrugarh, Assam, India, 786004}
		
  \author{Kalyan Malakar\orcidlink{0009-0002-5134-1553}}%
\email{kalyanmalakar349@gmail.com}
\affiliation{Department of Physics, Dibrugarh University, Dibrugarh, Assam, India, 786004}
\affiliation{Department of Physics, Silapathar College, Dhemaji, Assam, India, 787059}

	\author{Mrinnoy M. Gohain\orcidlink{0000-0002-1097-2124}}
	\email{mrinmoygohain19@gmail.com}
	\affiliation{%
		Department of Physics, Dibrugarh University, Dibrugarh, Assam, India, 786004}
	
	\author{Kalyan Bhuyan\orcidlink{0000-0002-8896-7691}}%
	\email{kalyanbhuyan@dibru.ac.in}
	\affiliation{%
		Department of Physics, Dibrugarh University, Dibrugarh, Assam, India, 786004}%
	\affiliation{Theoretical Physics Divison, Centre for Atmospheric Studies, Dibrugarh University, Dibrugarh, Assam, India 786004}

	
\begin{abstract}
In this work, we perform reconstruction of \( f(Q) \) gravity inspired by the New Agegraphic Dark Energy (NADE) model, aiming to account for the Universe's late time acceleration without invoking a cosmological constant. Utilizing a power law scale factor \( a(t) = a_0 t^h \), we derive an analytic form for \( f(Q) \) based on a correspondence with NADE, where the conformal time serves as the infrared cutoff. The resulting model naturally recovers General Relativity in the limit and exhibits a geometrically motivated dark energy component. We constrain the model parameters using recent Baryon Acoustic Oscillation (BAO) data from DESI DR2 BAO and previous BAO observations through the Markov Chain Monte Carlo (MCMC) analysis. The reconstructed Hubble parameter \( H(z) \) demonstrates excellent agreement with observational data, achieving high \( R^2 \) values and low \(\chi^2_{\min}\), AIC, and BIC scores, outperforming the standard \( \Lambda \)CDM model. Further, we investigate the cosmological evolution using the deceleration parameter \( q(z) \), effective equation of state \( \omega_{\mathrm{eff}}(z) \), and Om diagnostics. The model exhibits a clear transition from deceleration to acceleration with a present value \( q(0) \in \left[-0.5879, -0.3333\right] \) and transition redshift $z_{\mathrm{tr}} \sim 0.5209-0.8126$, while maintaining \( -1 < \omega_{\mathrm{eff}}(z) < -1/3 \), indicating quintessence like behavior. Om diagnostics consistently show a negative slope, further confirming deviation from \( \Lambda \)CDM. Energy condition analysis reveals that WEC, DEC, and NEC are satisfied, while SEC is violated only at low redshifts which is consistent with cosmic acceleration. Overall, the reconstructed \( f(Q) \) model provides a viable, observationally consistent, and theoretically motivated alternative to standard dark energy scenarios.
\end{abstract}

	
	\maketitle

\section{Introduction}\label{sec1}
The late-time acceleration of the universe has remained a central focus of cosmological research over the past two decades. This phenomenon has been firmly confirmed by several landmark observations, including Type Ia Supernovae (SNeIa) \cite{Riess1998, Perlmutter1999, Astier2006}, measurements of the Cosmic Microwave Background Radiation (CMBR) \cite{Spergel2003}, and data from large-scale structure (LSS) surveys \cite{Tegmark2004, Cole2005, Eisenstein2005}. Despite the observational success, the theoretical understanding of this accelerated cosmic expansion continues to pose significant challenges. General Relativity (GR), widely recognized as the most successful theory of gravitation, has effectively described the large-scale structure and evolution of the universe for over a century. Within the cosmological framework, the Friedmann–Lemaître–Robertson–Walker (FLRW) metric, along with appropriate matter content, provides viable solutions for the evolution of the cosmic scale factor \( a(t) \), capturing the universe’s dynamic behavior. There are two distinct epochs of accelerated expansion that the universe must undergo to align with observational data: an early-time inflationary phase and a late-time accelerated expansion. The former is often modeled through the introduction of a scalar field into the Einstein-Hilbert action, while the latter is addressed by incorporating a cosmological constant into the Einstein field equations. This straightforward framework forms the basis of the standard cosmological model, the so-called \(\Lambda\)CDM model, which remains consistent with nearly all available observational data. However, the model is not without its theoretical shortcomings. Chief among these is the cosmological constant problem, where the observed value of vacuum energy is many orders of magnitude smaller than what is predicted by quantum field theory \cite{Weinberg1989, Martin2012}. Furthermore, the fine-tuning and coincidence problems inherent to the \(\Lambda\)CDM model along with motivate the search for alternative explanations, either in the form of dynamic dark energy models or generalizations of Einstein’s theory of gravity.\\
Recently, a novel class of dark energy models has emerged based on quantum gravitational considerations, among which the \textit{Agegraphic Dark Energy} (ADE) model, originally proposed by Cai \cite{Cai2007}, stands out. Inspired by the Karolyhazy uncertainty relation and quantum fluctuations of spacetime, the ADE model assumes that the dark energy density originates from spacetime and matter field fluctuations. According to Karolyhazy et al. \cite{Karolyhazy1966}, the uncertainty in measuring a time interval \( t \) in Minkowski spacetime is given by \( \delta t \sim t_P^{2/3} t^{1/3} \), where \( t_P \) is the reduced Planck time. Building upon this, Maziashvili \cite{Maziashvili2007a, Maziashvili2007b} argued that the corresponding energy density of metric fluctuations takes the form \( \rho_{\Lambda} \sim M_P^2/t^2 \), where \( M_P \) is the reduced Planck mass. Using these quantum considerations, the ADE model was formulated to account for the accelerating expansion of the Universe. However, the original ADE model struggles to accommodate the matter-dominated epoch. To resolve this issue, Wei and Cai \cite{Wei2007} proposed the \textit{New Agegraphic Dark Energy} (NADE) model, in which the conformal time is chosen as the infrared cutoff instead of the age of the universe. This adjustment not only preserves the quantum origin of dark energy but also allows the model to naturally address the cosmic coincidence problem \cite{Wei2008}. The ADE and NADE models have since attracted considerable interest and have been extensively studied across a wide range of cosmological contexts \cite{Sheykhi2009, Karami2010, Karami2011a, Karami2011b}. In parallel, an alternative explanation for the observed cosmic acceleration lies in the realm of \textit{modified theories of gravity}, where the gravitational sector is extended to incorporate geometric corrections rather than invoking additional energy components. These theories can provide consistent cosmological dynamics and agree with current observations, sometimes even without the need for dark energy \cite{Capozziello2006, Nojiri2006recon, Sotiriou2010}. Among them, the recently proposed \textit{\( f(Q) \) gravity} models have gained significant attention. Unlike curvature-based \( f(R) \) or torsion-based \( f(T) \) models, \( f(Q) \) gravity is formulated in the framework of symmetric teleparallel geometry, where the gravitational effects arise solely from the non-metricity scalar \( Q \) \cite{Jimenez2018, Mandal2020}. This approach not only results in second-order field equations but also avoids the issues of local Lorentz violation and antisymmetric components typical of \( f(T) \) gravity.\\
One promising approach to addressing the cosmological constant problem and modeling late-time acceleration in modified gravity theories is the \textit{cosmological reconstruction method}. The primary aim of reconstruction is to derive a suitable form of the gravitational action that reproduces a known or observationally favored expansion history, such as that of the $\Lambda$CDM model. This method allows researchers to bypass direct assumptions about the functional form of the gravity theory and instead infer it based on cosmic evolution. However, the mathematical complexity of the field equations in modified gravity often prevents obtaining exact analytical solutions, and even numerical solutions can be computationally intensive and model-dependent. The reconstruction strategy, based on inverting the cosmological dynamics for a flat FLRW background, provides an effective workaround. It offers a systematic way to determine which class of modified theories—such as $f(R)$, $f(G)$, $f(Q)$, or $f(R,T)$—can mimic the desired cosmological behavior. Several important contributions exist in this domain. For instance, Nojiri et al. \cite{Nojiri2006recon} developed a method for reconstructing $f(R)$ gravity using the e-folding number as a key variable. Dunsby et al. \cite{Dunsby2010} showed that recovering the full $\Lambda$CDM dynamics in $f(R)$ gravity may require an additional effective matter component. Goheer et al. \cite{Goheer2009} demonstrated that only specific classes of $f(G)$ models can yield exact power-law solutions. More recently, a variety of cosmological scenarios have been explored using reconstruction techniques in theories like $f(\tau, T)$ gravity \cite{Baffou2017} and $f(Q)$ gravity \cite{Saha2025}. This line of research opens the door to constructing phenomenologically viable modified gravity models that are not only compatible with observational data but also potentially free from the theoretical pitfalls of $\Lambda$CDM.\\
In this work, we focus on investigating a cosmological reconstruction scheme within the framework of $f(Q)$ gravity, employing a direct correspondence with the New Agegraphic Dark Energy (NADE) model. Given that NADE—and more broadly, holographic dark energy (HDE)—has its theoretical foundation in black hole thermodynamics and the holographic principle, it naturally incorporates quantum gravitational effects into cosmological modeling. This makes it an ideal candidate for reconstructing modified gravity models that aim to unify early- and late-time cosmological dynamics. Previous studies have demonstrated that reconstruction approaches using various forms of dark energy—such as HDE, NADE, and their entropy-corrected extensions—can successfully yield viable modified gravity functions \cite{Karami2011a, Saha2025, Nojiri2006recon}. In particular, when combined with frameworks like $f(R)$ or $f(T)$ gravity, these methods have produced models consistent with observational constraints and capable of reproducing essential cosmic epochs. Motivated by these findings, we aim to extend this approach to $f(Q)$ gravity, which offers a geometrically rich and dynamically versatile alternative to curvature- and torsion-based theories. Our objective is to reconstruct the functional form of $f(Q)$ gravity using NADE energy density as input, and to examine the cosmological consequences of the resulting models—especially their ability to account for late-time acceleration. The quantum origin of NADE further enhances the theoretical interest of this study, potentially providing insight into the intersection of modified gravity and quantum cosmology.\\
The structure of this paper is as follows: Sec. \ref{s2} provides a brief mathematical formalism of $f(Q)$ gravity and the cosmological framework in a flat FLRW Universe, Sec. \ref{s3} describes the reconstruction for $f(Q)$ gravity using NADE. Sec. \ref{s4} discusses the observational data and methodology used in the study to constraint the model. Sec \ref{s5} discuses the constraints on the models and the cosmological implications it represents. Finally, Sec. \ref{s6} provides a conclusive summary of the study.
\section{$f(Q) Gravity$}\label{s2}
In this section, we briefly present the theoretical structure of $f(Q)$ gravity, where the gravitational interaction is attributed to the non-metricity scalar $Q$, instead of the curvature scalar $R$ or the torsion scalar $T$. That is in $f(Q)$ gravity, the connection is assumed to be curvature- and torsion-free, but with non-vanishing non-metricity \cite{Jimenez2018}:
\begin{equation}
Q^\gamma_{\mu\nu} \equiv -\nabla_\gamma g_{\mu\nu}.
\end{equation}
This non-metricity quantifies how the length of a vector changes under parallel transport. The affine connection can be decomposed as:
\begin{equation}
\Upsilon^\gamma_{\mu\nu} = \Gamma^\gamma_{\mu\nu} + L^\gamma_{\mu\nu},
\end{equation}
where $\Gamma^\gamma_{\mu\nu}$ is the Levi-Civita connection, and $L^\gamma_{\mu\nu}$ is the disformation tensor constructed from the non-metricity tensor. The non-metricity scalar $Q$ is constructed from contractions of the non-metricity tensor and a superpotential tensor $P^\gamma_{\mu\nu}$, defined as:
\begin{equation}
Q = -Q^\gamma_{\mu\nu} P^\mu_\gamma{}^\nu.
\end{equation}
The action for $f(Q)$ gravity in the presence of matter fields is given by \cite{Jimenez2018}:
\begin{equation}
S = \int d^4x \sqrt{-g} \left[ -\frac{1}{2} f(Q) + \mathcal{L}_m \right],
\end{equation}
where $f(Q)$ is a general function of the non-metricity scalar and $\mathcal{L}_m$ denotes the matter Lagrangian. Varying the action with respect to the metric yields the modified gravitational field equations:
\begin{equation}
2 \nabla_\gamma \left( \sqrt{-g} f_Q P^\gamma_{\mu\nu} \right) + \sqrt{-g} \left[ \frac{1}{2} f g_{\mu\nu} + f_Q \left(P_{\nu\rho\sigma} Q_\mu{}^{\rho\sigma} - 2 P_{\rho\sigma\mu} Q^\rho{}_{\nu}{}^\sigma \right) \right] = \sqrt{-g} T_{\mu\nu},
\end{equation}
where $f_Q = df/dQ$ and $T_{\mu\nu}$ is the energy-momentum tensor. In the coincident gauge, where the connection vanishes ($\Upsilon^\gamma_{\mu\nu} = 0$), the spacetime becomes flat and simplifies calculations. In this gauge, the non-metricity tensor becomes $Q^\gamma_{\mu\nu} = -\partial_\gamma g_{\mu\nu}$, making the covariant derivatives reduce to partial derivatives.\\
To explore cosmological implications, we assumes a spatially flat Friedmann–Lemaître–Robertson–Walker (FLRW) metric:
\begin{equation}
ds^2 = -dt^2 + a^2(t)\left( dx^2 + dy^2 + dz^2 \right),
\end{equation}
where $a(t)$ is the scale factor. For this metric, the non-metricity scalar becomes:
\begin{equation}
Q = 6H^2,
\end{equation}
with $H = \dot{a}/a$ as the Hubble parameter. The energy-momentum tensor, $T_{\mu\nu} = (\rho + p)u_\mu u_\nu + p g_{\mu\nu}$, is taken into consideration, where $u^\mu = (-1,0,0,0).$ represents the four-velocity of the fluid that satisfies the condition $u^\mu u_\mu = -1$,$\rho$ and $p$ are the pressure and energy density, respectively. With the aforementioned metric and a perfect fluid energy-momentum tensor, the modified Friedmann equations in $f(Q)$ gravity become:
\begin{equation}
3H^2 = \frac{1}{2f_Q} \left( \rho + \frac{f}{2} \right), \qquad \dot{H} + 3H^2 + \frac{\dot{f}_Q}{f_Q} H = \frac{1}{2f_Q} \left( -p + \frac{f}{2} \right).
\end{equation}
Assuming $8\pi G = c = 1$, the overhead dot (.) represents the derivative with regard to time.  These equations explain how the cosmological constant, spatial curvature, and the energy density of matter and radiation affect the universe's expansion. The usual GR equations are restored if we put $f(Q) = Q$. However, when we insert $f(Q)=Q+F(Q)$, the field equations change to:
\begin{equation}
3H^2 = \rho + F - 2QF_Q,
\end{equation}
\begin{equation}
\left(2QF_{QQ} + F_Q + 1\right) \dot{H} + \frac{1}{4}\left(Q + 2QF_Q - F\right) = -2p,
\end{equation}
where \( F_Q = \frac{dF}{dQ} \) and \( F_{QQ} = \frac{d^2F}{dQ^2} \). In the context of isotropic and homogeneous spatially flat FLRW cosmologies these equation can be further defined as:
\begin{equation}
3H^2 = \rho_m + \rho_r + \rho_{de}, \qquad 2\dot{H} + 3H^2 = -(p_m + p_r + p_{de}).
\end{equation}
Here, $\rho_r$ and $\rho_m$ represent the energy densities of the radiation and matter components, with $p_m$ and $p_r$ indicating the pressure associated with matter and radiation components respectively. Along with the effective dark energy density $\rho_{de}$ and pressure $p_{de}$ arising from the geometric contribution can be written as:
\begin{equation}
\rho_{de} = \frac{F}{2} - Q F_Q,
\label{e12}
\end{equation}
\begin{equation}
p_{de} = 2 \dot{H} \left(2Q F_{QQ} + F_Q\right) - \rho_{de},
\end{equation}
Considering non-interacting radiation, matter, and geometric dark energy components, their energy conservation equations are:
\begin{equation}
\dot{\rho}_r + 4H\rho_r = 0, \qquad \dot{\rho}_m + 3H\rho_m = 0, \qquad \dot{\rho}_{de} + 3H(1 + \omega_{de})\rho_{de} = 0.
\end{equation}
where, $\omega_{de}=\frac{p_{de}}{\rho_{de}}$. These can help to derive standard redshift evolution:
\begin{equation}
\rho_r(z) = \rho_{r0}(1 + z)^4, \qquad \rho_m(z) = \rho_{m0}(1 + z)^3.
\end{equation}
where $\rho_{r0}$ and $\rho_{m0}$ denote the present values of $\rho_{r}$ and $\rho_{r}$.

\section{Reconstruction of \boldmath{$f(Q)$} Gravity from NADE}
\label{s3}
In this section, we present the reconstruction scheme that we adopt to determine the functional form of $f(Q)$ gravity using the New Agegraphic Dark Energy (NADE) model as a guiding principle. The NADE model, originally proposed by Wei and Cai \cite{Wei2007}, is based on the Karolyhazy uncertainty relation and quantum fluctuations of spacetime, and uses the conformal time $\eta$ as the infrared (IR) cutoff. The energy density of NADE is given by:
\begin{equation}
\rho_{\text{NADE}} = \frac{3 n^2 M_P^2}{\eta^2},
\label{e16}
\end{equation}
where $n$ is a dimensionless parameter, $M_P^2 = 1/\kappa^2$ is the reduced Planck mass squared, and $\eta$ is the conformal time defined as:
\begin{equation}
\eta = \int \frac{dt'}{a(t')} = \int \frac{da'}{a'^2 H(a')}.
\label{e17}
\end{equation}
Combining Eq. (\ref{e16}) and Eq. (\ref{e12}) we obtain a differential equation for $f(Q)$:
\begin{equation}
 F(Q) - 2Q F_Q = 6 \frac{n^2}{\eta^2}.
\label{eq:recon_diff}
\end{equation}
where, we have taken $\kappa^2 = 1$. This is a first-order linear differential equation in $F(Q)$. Solving this equation requires expressing the conformal time $\eta$ in terms of $Q$. This can be done by choosing a specific form of the scale factor $a(t)$ (or equivalently $H(t)$), from which we compute $H(t)$, $Q(t)$, and $\eta(t)$. Eliminating the time parameter then gives $\eta = \eta(Q)$, allowing us to fully reconstruct $f(Q)$ from Eq. \eqref{eq:recon_diff}.\\ 
Literature shows there are different classes of scale factors which usually are consider to illustrate the
accelerating universe in various modified gravities \cite{f1}. And, one of those scale factors is given as:
\begin{equation}
a(t) = a_0 t^h, \quad h > 0,
\label{e19}
\end{equation}
for which one can obtain the following cosmological quantities:
\begin{equation}
H = \frac{h}{t} , \quad \dot{H} = -\frac{H^2}{h},
\end{equation}
Now, using the scale factor defined by Eq. (\ref{e19}) with Eq. (\ref{e17}) we obtain the conformal time as:
\begin{equation}
\eta = \int_0^t \frac{dt'}{a(t')} = \frac{t^{1-h}}{a_0(1-h)} = \frac{\left(\frac{6 h^2}{Q}\right)^{\frac{1-h}{2}}}{a_0 (1-h)}
\label{e21}
\end{equation}
Using Eq. (\ref{e21}) with Eq. (\ref{eq:recon_diff}) we obtain:
\begin{equation}
 F(Q) - 2Q F_Q= 6 \frac{n^2 \left(a_0-a_0 h\right){}^2}{\left(\frac{6 h^2}{Q}\right)^{1-h}}.
\end{equation}
Solving this differential equation we obtain the reconstructed $f(Q)$ gravity as:
\begin{equation}
f(Q)=Q+F(Q), \quad
F(Q) = \sqrt{Q} \left(\frac{a_0^2 6^h (h-1)^2 n^2 \left(\frac{h}{\sqrt{Q}}\right)^{2 h-1}}{h (2 h-1)}+C_1\right)
\label{e23}
\end{equation}
Here, $C_1$ is a some arbitrary constant. For the reconstructed $F(Q)$ gravity model it can be seen that when $h \to 1$ and $C_1 \to 0$, we get $f(Q) \to Q$ that is the standard GR gets recovered. In the following section we constraint the parameters using observational data and investigate the cosmological implication it provides.
\section{Observational data and Methodology for Constraints}\label{s4}
For obtaining the constrain on the parameters of the reconstructed $f(Q)$ gravity model we employ a combination of Baryon Acoustic Oscillation (BAO) data from both pre-DESI or previous BAO (from observations such as SDSS and WiggleZ) and recent DESI DR2 BAO releases for the Hubble parameter $H(z)$. Other cosmological probes such as Type Ia supernovae  or cosmic chronometers can also be used for this purpose but for simplicity of the task we focus only on the BAO data. We analyze the data through a Markov Chain Monte Carlo (MCMC) framework to obtain robust constraints on the cosmological and model parameters. Under the light of which then the cosmological scenarios are studied using key cosmological indicators such as the deceleration parameter \( q(z) \), effective equation of state \( w_{\text{eff}}(z) \), $om$ diagnostics, and finally the energy conditions.\\
Literature shows using the reconstructed \( f(Q) \) model, the modified Friedmann equation can be recast as:
\begin{equation}
(\frac{H(z)}{H_0})^{2} = \Omega_{r0} (1 + z)^4 + \Omega_{m0} (1 + z)^3 + \frac{1}{3H_0^2} \left[ \frac{1}{2}F(Q(z)) - Q(z) F_Q(Q(z)) \right],
\label{e29}
\end{equation}
Here the current values of the radiation and matter density parameters are represented by \(\Omega_{r0}\) and \(\Omega_{m0}\), respectively, whereas the present value of the Hubble parameter is $H_0$. This formulation allows for direct comparison with observational Hubble data and facilitates parameter estimation via standard statistical techniques. Further ones $H(z)$ is obtained we can obtain the deceleration parameter, given as:
\begin{equation}
    q(z) = -1 - \frac{\dot{H}}{H^2} = -1 + (1+z) \frac{1}{H(z)} \frac{dH}{dz},
    \label{eq}
\end{equation}
which indicates the acceleration or deceleration of the Universe. A negative value of $q(z)$ corresponds to accelerated expansion. Also, we can obtain the effective EoS, given as:
\begin{equation}
\omega_{eff}(z) = -1 + \frac{2(1+z)}{3H(z)}\frac{dH}{dz},
 \label{ew}
\end{equation}
which provides an effective fluid description of the Universe's dynamics, incorporating both matter and dark energy effects. These quantities allow us to examine whether the model predicts a transition from deceleration to acceleration and to test whether the effective equation of state crosses the phantom divide ($w_{\text{eff}} < -1$). Apart from that we can also have the $Om(z)$ diagnostic, it is defined as:
\begin{equation}
Om(z) = \frac{H^2(z)/H_0^2 - 1}{(1 + z)^3 - 1}.
\label{eo}
\end{equation}
It is a pure geometrical tool constructed from the Hubble parameter, allowing for discrimination among dark energy models without requiring knowledge of the equation of state.
\begin{table}[h!]
\centering
\begin{tabular}{|ccc|c||ccc|c|}
\hline
\multicolumn{4}{|c||}{\textbf{DESI}} & \multicolumn{4}{c|}{\textbf{P-BAO}} \\
\hline
$z$ & $H(z)$ & $\sigma_H$ & Ref & $z$ & $H(z)$ & $\sigma_H$ & Ref \\
\hline
0.51 & 97.21 & 2.83 & \cite{ref98} & 0.24 & 79.69 & 2.99 & \cite{d113} \\
0.71 & 101.57 & 3.04 & \cite{ref98} & 0.30 & 81.70 & 6.22 & \cite{d114} \\
0.93 & 114.07 & 2.24 & \cite{ref98} & 0.31 & 78.17 & 6.74 & \cite{d115} \\
1.32 & 147.58 & 4.49 & \cite{ref98} & 0.34 & 83.17 & 6.74 & \cite{d113} \\
2.33 & 239.38 & 4.80 & \cite{ref98} & 0.35 & 82.70 & 8.40 & \cite{d116} \\
 &  &  &  & 0.36 & 79.93 & 3.39 & \cite{d115} \\
 &  &  &  & 0.38 & 81.50 & 1.90 & \cite{d5} \\
 &  &  &  & 0.40 & 82.04 & 2.03 & \cite{d115} \\
 &  &  &  & 0.43 & 86.45 & 3.68 & \cite{d113} \\
 &  &  &  & 0.44 & 82.60 & 7.80 & \cite{d74} \\
 &  &  &  & 0.44 & 84.81 & 1.83 & \cite{d115} \\
 &  &  &  & 0.48 & 87.79 & 2.03 & \cite{d115} \\
 &  &  &  & 0.56 & 93.33 & 2.32 & \cite{d115} \\
 &  &  &  & 0.57 & 87.60 & 7.80 & \cite{d10} \\
 &  &  &  & 0.57 & 96.80 & 3.40 & \cite{d117} \\
 &  &  &  & 0.59 & 98.48 & 3.19 & \cite{d115} \\
 &  &  &  & 0.60 & 87.90 & 6.10 & \cite{d74} \\
 &  &  &  & 0.61 & 97.30 & 2.10 & \cite{d5} \\
 &  &  &  & 0.64 & 98.82 & 2.99 & \cite{d115} \\
 &  &  &  & 0.978 & 113.72 & 14.63 & \cite{d118} \\
 &  &  &  & 1.23 & 131.44 & 12.42 & \cite{d118} \\
 &  &  &  & 1.48 & 153.81 & 6.39 & \cite{d79} \\
 &  &  &  & 1.526 & 148.11 & 12.71 & \cite{d118} \\
 &  &  &  & 1.944 & 172.63 & 14.79 & \cite{d118} \\
 &  &  &  & 2.30 & 224.00 & 8.00 & \cite{d119} \\
 &  &  &  & 2.36 & 226.00 & 8.00 & \cite{d120} \\
 &  &  &  & 2.40 & 227.80 & 5.61 & \cite{d121} \\
\hline
\end{tabular}
\caption{Observed Hubble parameter $H(z)$ (in units of $km s^{-1} Mpc^{-1}$) and their uncertainties at redshift $z$ form the DESI and P-BAO datasets.}
\label{table:4}
\end{table}\\
To find the mean values of the best fitting parameters for the model, we employ the chi-squared minimization method using the following chi-squared function:
\begin{equation}
\chi^2_H(p_1,p_2,...) = \sum_{i=1}^{N} \frac{\left[ H_{th}(z_i,p_1,p_2,..) - H_{obs}(z_i) \right]^2}{\sigma_H(z_i)^2},
\label{32}
\end{equation}
where $H_{th}$ is the theoretical value of the Hubble parameter, $H_{obs}$ is the observed value, and $\sigma_H(z_i)$ is the standard error in the observed value of $H(z_i)$ at redshift $z_i$. We conduct the MCMC analysis in three stages: using the DESI dataset alone, then with the previous BAO dataset, and finally with a combination of both. The outcomes are compared with those from the standard \(\Lambda\)CDM model using statistical indicators such as the coefficient of determination \( R^2 \), the minimum chi-squared value \( \chi^2_{\min} \), the Akaike Information Criterion (AIC), and the Bayesian Information Criterion (BIC) (see Refs.~\cite{refS66}, \cite{refS68}, \cite{refS64} for details). A model with lower AIC, BIC, and \( \chi^2_{\min} \) values is statistically preferable because it shows a better balance between goodness of fit and complexity. Among these, \( R^2 \) and \( \chi^2_{\min} \) assess fit quality without accounting for the number of free parameters. In contrast, AIC and BIC penalize models with greater complexity, though BIC applies a stronger penalty, especially with larger datasets. To evaluate the relative performance of models, we compute the differences in AIC and BIC as \( \Delta X = \Delta \text{AIC} \) or \( \Delta X = \Delta \text{BIC} \). These differences are interpreted as follows:
\begin{itemize}
\item \textbf{$0 \leq \Delta X \leq 2$}:The evidence is \textit{weak} and it is impossible to judge whether model is superior.
\item \textbf{$2 < \Delta X \leq 6$}: Evidence is \textit{positive} in support of the model with the lower value. 
\item \textbf{$6 < \Delta X \leq 10$}: Evidence is considered to be \textit{strong}.
\end{itemize}
These statistical diagnostics serve as essential tools for model comparison, ensuring both accuracy and parsimony in cosmological inference.
\section{Constraints and Cosmological Implications}\label{s5}
\begin{figure*}[htb]
\centerline{
\includegraphics[width=.85\textwidth]{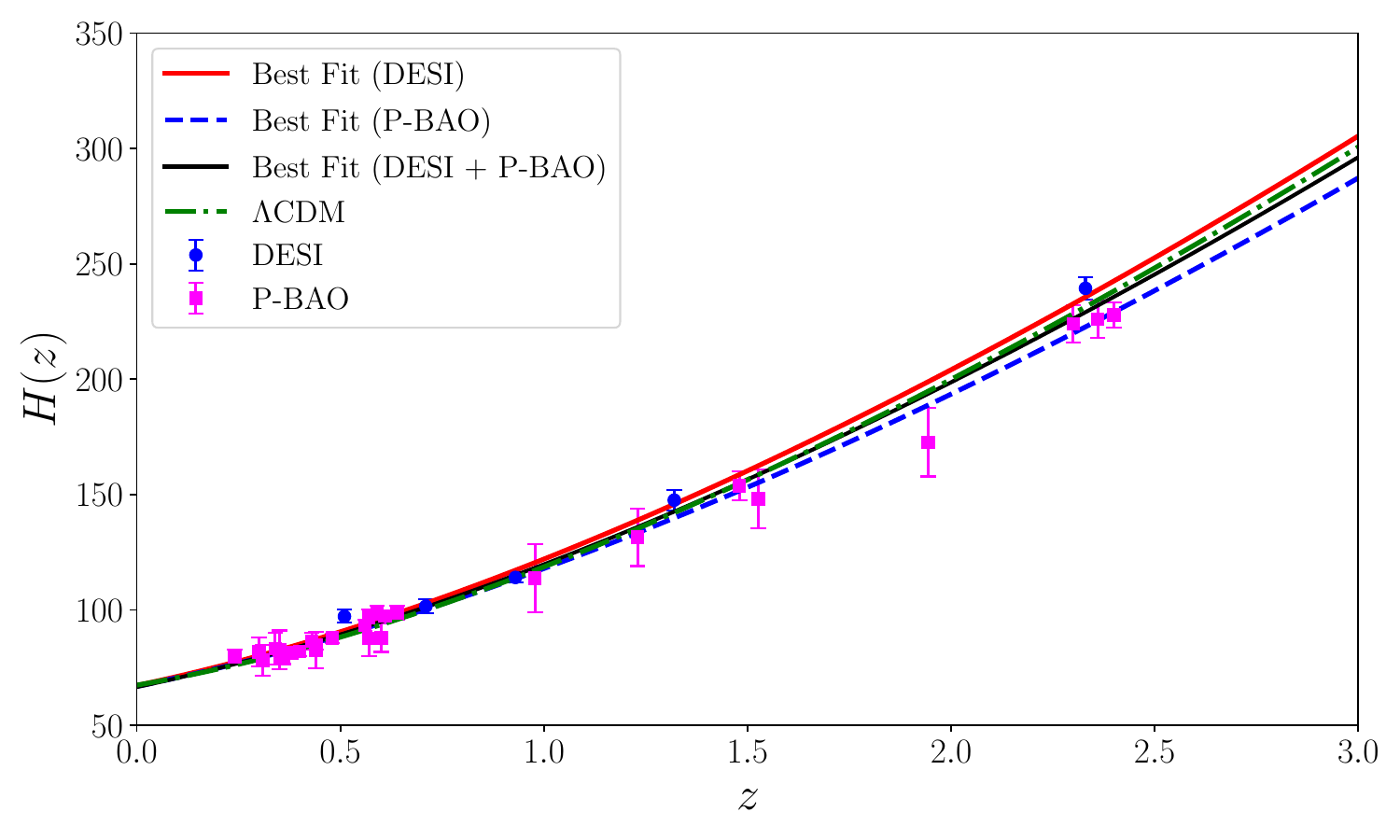}}
\caption{Plot of $H(z)$ vs $z$ for the best fit of the model parameters against the observational data. Here, a comparison is made with the $\Lambda$CDM model.}
\label{f1}
\end{figure*}

Using the observational datasets and methodology as mention in the above section, we investigate the constraints and cosmological implications for the reconstructed $f(Q)$ gravity model. Here, using the scale factor given by Eq. (\ref{e19}), we can obtain the Hubble parameter as:
\begin{equation}
H(z) = H_0 (1 + z)^{1/h}
\end{equation}
Now, as we know $Q=6H^{2}$ this implies $Q=(H_0(1+z)^{1/h})^2$, using it with Eqs. (\ref{e23}) and (\ref{e29}) we obtain:
\begin{equation}
\frac{H(z)^2}{H_0^2}= (1 + z)^3 \, \Omega_{m0} + (1 + z)^4 \, \Omega_{r0} +
\frac{
\sqrt{H_0^2 (1 + z)^{2/h}} \left(
\frac{ \sqrt{6} \, a_0^2 (1 - h)^2 n^2 \left( \frac{h}{\sqrt{H_0^2 (1 + z)^{2/h}}} \right)^{-1 + 2h} }{ h (2h - 1) } + C_1
\right)
}{
\sqrt{6} \, H_0^2
}
\label{e33}
\end{equation}
Here, we can constraint $C_1$ by the condition the $\frac{H(z)}{H_0}=1$ when $z=0$, from which we get:
\begin{equation}
C_1=\sqrt{6} \, \sqrt{H_0^2} \left( 
1 - \frac{ a_0^2 (1 - h)^2 \left( \dfrac{h}{\sqrt{H_0^2}} \right)^{-1 + 2h} n^2 }{ h (2h - 1) \sqrt{H_0^2} } 
- \Omega_{m0} - \Omega_{r0}
\right)
\label{e34}
\end{equation} 
Using Eq. (\ref{e34}) in Eq. (\ref{e33}) along with the observation datasets as  described previously, we obtain a constraint on the model parameters. Results for which are shown in Fig. (\ref{f1}) along with Tables (\ref{tab1}) and (\ref{tabr1}). Results from Table (\ref{tabr1}) shows that the reconstructed $f(Q)$ model consistently achieves higher $R^2$ values and lower $\chi^2_{\min}$, AIC, and BIC scores across all datasets in comparison to the standard $\Lambda$CDM model. The DESI dataset, in particular, shows a strong preference for $f(Q)$ with $\Delta$AIC = 2.03 and $\Delta$BIC = 2.82, indicating positive evidence in favor of the reconstructed model. The P-BAO dataset also favors $f(Q)$ with moderate improvements, while the combined dataset yields $\Delta$AIC = 8.83 and $\Delta$BIC = 5.39, corresponding to strong support. These results demonstrate that the reconstructed model provides a superior fit without with respect to the observational data without requiring the presence of a cosmological constant.\\
Finally using Eq. (\ref{e33}) along with Eqs. (\ref{e34}), (\ref{eq}), (\ref{ew}), and (\ref{eo}), we can obtain the deceleration parameter, effective EoS, and $Om$ diagnostics for the for the model. We avoid writing the explicit form of these quantiles because of their complicated forms, and hence rather provide a graphical illustration of them for the best values of the model parameters as shown in Figs. (\ref{f21}), (\ref{f22}) and (\ref{f23}). Along with that we also evaluate the respective energy conditions also for the reconstructed $f(Q)$ gravity model, signifying better agreement with observations.

\begin{figure*}[htb]
\centerline{
\includegraphics[width=.85\textwidth]{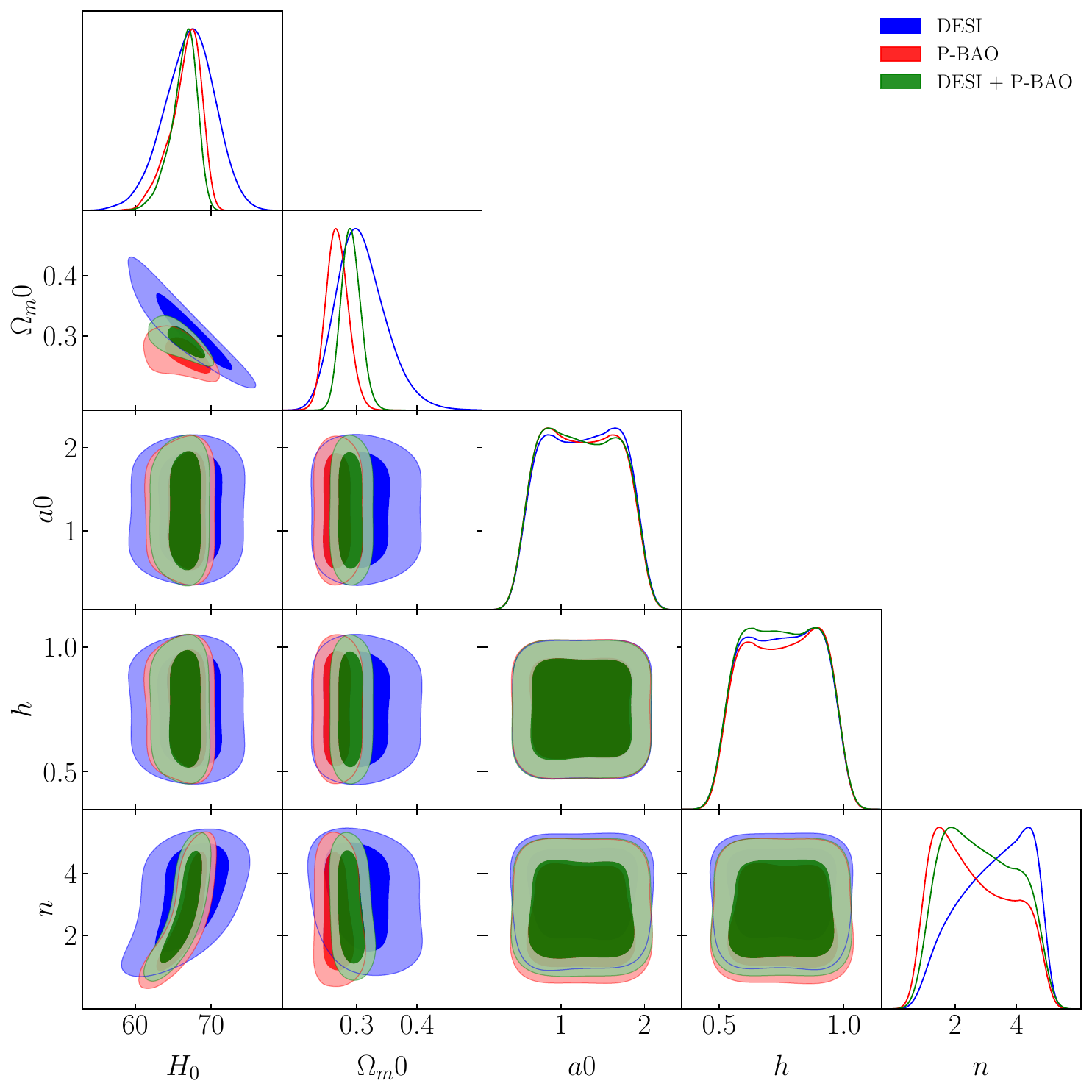}}
\caption{2-d contour sub-plot for the parameters $H_0$, $n$, $\omega_0$ and $\omega_1$ with 1-$\sigma$ and 2-$\sigma$ errors (showing the 68\% and 95\% c.l.) for $H(z)$ vs $z$.}
\label{figt1}
\end{figure*}

\begin{table}[htbp]
\centering
\renewcommand{\arraystretch}{1.2}
\begin{tabular}{lccccc}
\toprule
\textbf{Dataset} & \boldmath$H_0$ & \boldmath$\Omega_{m0}$ & \boldmath$a_0$ & \boldmath$h$ & \boldmath$n$ \\
\midrule
DESI & 
$67.17^{+3.24}_{-3.38}$ &
$0.306^{+0.044}_{-0.037}$ &
$1.251^{+0.510}_{-0.508}$ &
$0.751^{+0.170}_{-0.172}$ &
$3.401^{+1.119}_{-1.371}$ \\
P-BAO & 
$66.89^{+1.67}_{-2.43}$ &
$0.267^{+0.018}_{-0.017}$ &
$1.228^{+0.522}_{-0.500}$ &
$0.752^{+0.169}_{-0.171}$ &
$2.458^{+1.603}_{-1.160}$ \\
DESI + P-BAO & 
$66.59^{+1.43}_{-2.01}$ &
$0.290^{+0.017}_{-0.016}$ &
$1.241^{+0.514}_{-0.505}$ &
$0.757^{+0.167}_{-0.174}$ &
$2.808^{+1.434}_{-1.255}$ \\
\bottomrule
\end{tabular}
\caption{Best-fit values of model parameters from MCMC fitting of the reconstructed $f(Q)$ gravity model using different datasets. The values are quoted with 1$\sigma$ uncertainties.}
\label{tab1}
\end{table}

\begin{table*}[ht!]
\centering
\renewcommand{\arraystretch}{1.2}
\begin{tabular}{llcccccc}
\toprule
\textbf{Model} & \textbf{Dataset} & \textbf{$R^2$} & \textbf{$\chi^2_{\min}$} & \textbf{AIC} & \textbf{BIC} & \textbf{$\Delta$AIC} & \textbf{$\Delta$BIC} \\
\midrule
{\textbf{New Model}} 
  & DESI     & 0.9953 & 7.55  & 17.55 & 15.59 & 2.03  & 2.82 \\
  & P-BAO     & 0.9912 & 13.62 & 23.62 & 30.10 & 3.41  & 0.81 \\
  & DESI       & 0.9860 & 21.17 & 31.17 & 39.61 & 8.83  & 5.39. \\
\midrule
{\textbf{$\Lambda$CDM}} 
  & DESI     & 0.9878 & 13.58 & 19.58 & 18.41 & --    & --   \\
  & P-BAO     & 0.9830 & 21.03 & 27.03 & 30.91 & --    & --   \\
  & DESI + P-BAO   & 0.9846 & 34.60 & 40.60 & 45.00 & --    & --   \\
\bottomrule
\end{tabular}
\caption{Comparison of the new $f(Q)$ cosmological model with $\Lambda$CDM using three observational datasets. $\Delta$AIC and $\Delta$BIC are calculated as $\Lambda$CDM $-$ Model. A positive $\Delta$AIC/BIC indicates a statistical preference for the $f(Q)$ model.}
\label{tabr1}
\end{table*}

\begin{table}[htbp]
\centering
\renewcommand{\arraystretch}{1.2}
\begin{tabular}{lccc}
\toprule
Dataset & $q(0)$ & $w_{\text{eff}}(0)$ & $z_{\text{tr}}$ \\
\midrule
DESI         & $-0.4330^{+0.0997}_{-0.0676}$ & $-0.6220^{+0.0665}_{-0.0451}$ & $0.6473^{+0.1213}_{-0.1264}$ \\
P-BAO        & $-0.4480^{+0.1399}_{-0.0683}$ & $-0.6320^{+0.0932}_{-0.0455}$ & $0.7512^{+0.0614}_{-0.0834}$ \\
DESI+P-BAO   & $-0.4361^{+0.1069}_{-0.0531}$ & $-0.6241^{+0.0713}_{-0.0354}$ & $0.6806^{+0.0574}_{-0.0647}$ \\
\bottomrule
\end{tabular}
\caption{Key cosmological indicators at present and transition epoch from reconstructed $f(Q)$ gravity.}
\label{tabr2}
\end{table}
\subsection{Deceleration Parameter}
\begin{figure*}[htb]
\centerline{
\includegraphics[width=1\textwidth]{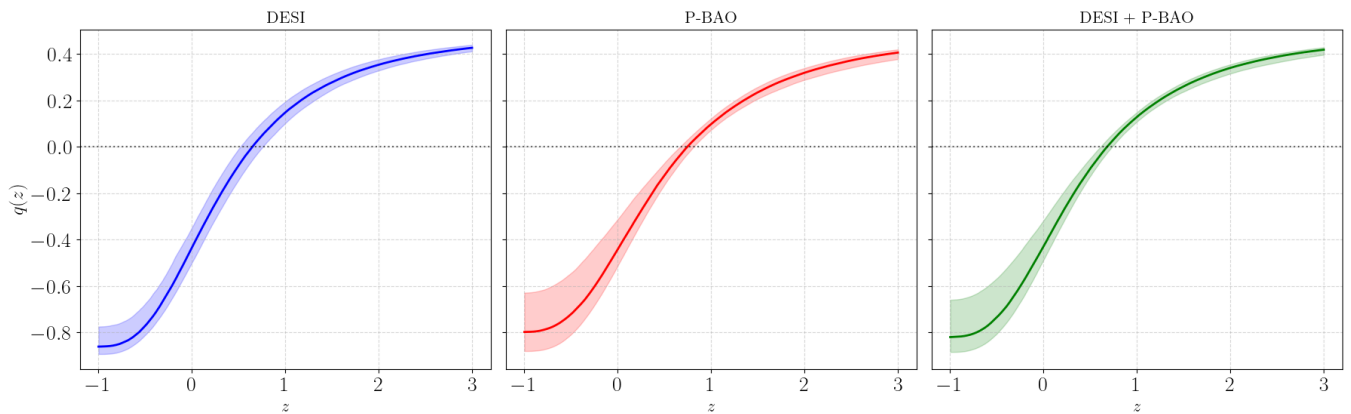}}
\caption{Evolution of deceleration parameter with redshift with best model parameters for the first model. The shaded regions here indicate the allowed region at 1$\sigma$ confidence level.}
\label{f21}
\end{figure*}
The deceleration parameter \( q(z) \) plays a central role in understanding the expansion history of the Universe. A positive $q$ implies deceleration (as in a matter- or radiation-dominated era), whereas a negative $q$ indicates cosmic acceleration driven by dark energy. The transition from deceleration to acceleration is considered a defining feature of late-time cosmic evolution, now well-supported by a wealth of observations. In our models, the evolution of $q(z)$ clearly demonstrates a transition from a decelerated to an accelerated phase. For our reconstructed $f(Q)$ model, we find that the present value of the deceleration parameter lies in the range $-1 < q(0) < 0$, consistent with an accelerating Universe. Details for which are illustrated by Fig. (\ref{f21}) and Table~\ref{tabr2}. More precisely the datasets yield consistent negative values for \( q(0) \), with:
$q(0) \in [-0.5879, -0.3333]$
confirming the current accelerated expansion of the Universe. The combined dataset (DESI + P-BAO) gives \( q(0) = -0.4361^{+0.1069}_{-0.0531} \), which is compatible with observational estimates from Planck \cite{planck2018}, and recent Hubble tension discussions \cite{verde2019tensions}. The exact transition redshift, $z_{\text{tr}}$, at which the Universe switches from deceleration to acceleration, for all the datasets fall within the range,
$z_{\rm tr} \in [0.5209, 0.8126]$ which agrees well with earlier studies estimating the acceleration onset near \( z \sim 0.6 - 0.8 \) \cite{farooq2013, xu2012}. The transition redshift values  Such dynamics supports the idea that the reconstructed $f(Q)$ model, effectively captures the late-time acceleration without invoking the need of a cosmological constant explicitly.
\subsection{Effective EoS}
\begin{figure*}[htb]
\centerline{
\includegraphics[width=1\textwidth]{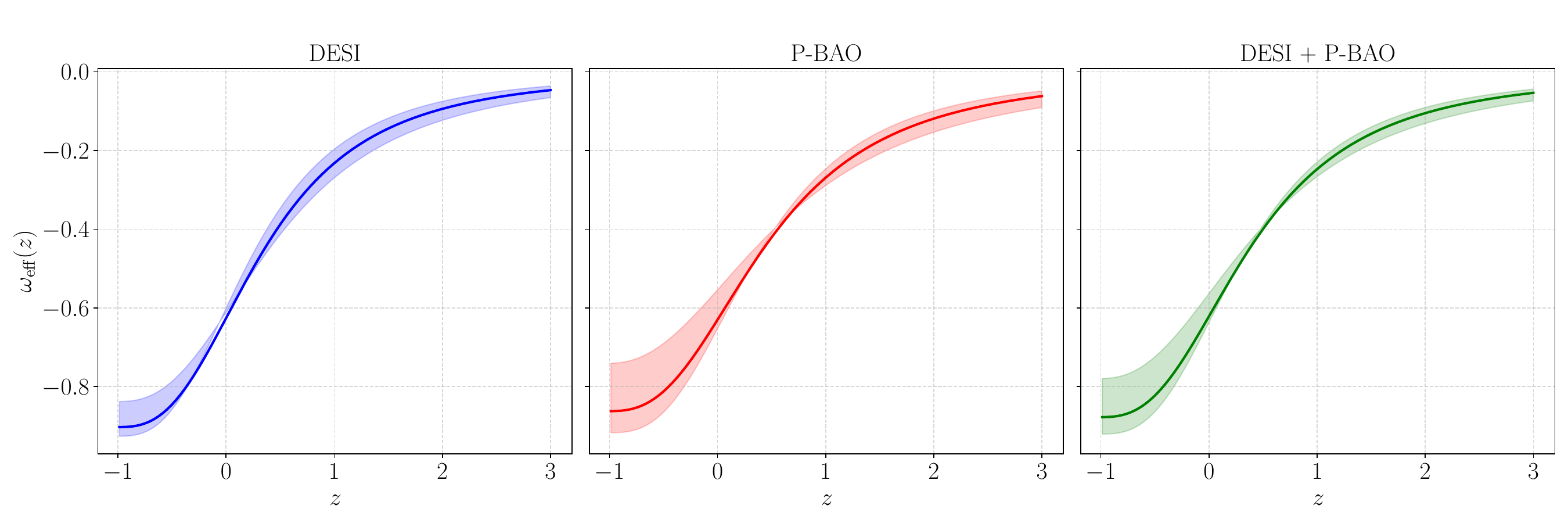}}
\caption{Evolution of effective EoS with redshift with best model parameters for the first model. The shaded regions here indicate the allowed region at 1$\sigma$ confidence level.}
\label{f22}
\end{figure*}
The effective equation of state (EoS) parameter, $\omega_{\rm eff}(z)$, offers a unified view of the cosmic fluid's dynamical evolution, encapsulating the contributions from matter, radiation, and geometric corrections arising from the reconstructed $f(Q)$ gravity model. For $\omega_{\rm eff}(z) < -1/3$, the Universe undergoes accelerated expansion, while $\omega_{\rm eff}(z) = -1$ represents the $\Lambda$CDM limit. If $\omega_{\rm eff}(z) < -1$, the Universe enters the phantom regime, and $-1 < \omega_{\rm eff}(z) \leq -1/3$ indicates quintessence-like behavior. Our reconstructed $f(Q)$ gravity model exhibits a clear transition from matter domination to a quintessence-like accelerated phase. The present-day effective EoS values derived from different datasets consistently fall within the range $-0.725 < \omega_{\rm eff}(0) < -0.586$, suggesting that the model supports a Universe currently in a quintessence-dominated accelerated epoch. Detailed results are shown in Fig. ( \ref{f22}) and Table \ref{tabr2}. From Table~\ref{tabr2}, we observe that the DESI dataset favours a slightly less negative value of $\omega_{\rm eff}(0)$, implying milder acceleration compared to other datasets. The combined DESI+P-BAO dataset yields a central value of $-0.6241$, which closely aligns with Planck 2018 estimates for dynamical dark energy \cite{planck2018}, and is marginally less negative than values observed in models predicting phantom evolution \cite{verde2019tensions}. The evolution of $\omega_{\rm eff}(z)$, shown in Fig.~\ref{f22}, indicates that at higher redshifts, the EoS approaches values closer to $0$, which is consistent with the expected matter-dominated era. As $z \to 0$, all datasets predict a transition into the acceleration phase without requiring a cosmological constant. These results strengthen the viability of the reconstructed $f(Q)$ gravity model as an effective geometrical alternative to $\Lambda$CDM.
\subsection{$Om$ Diagnostics}
\begin{figure*}[htb]
\centerline{
\includegraphics[width=1\textwidth]{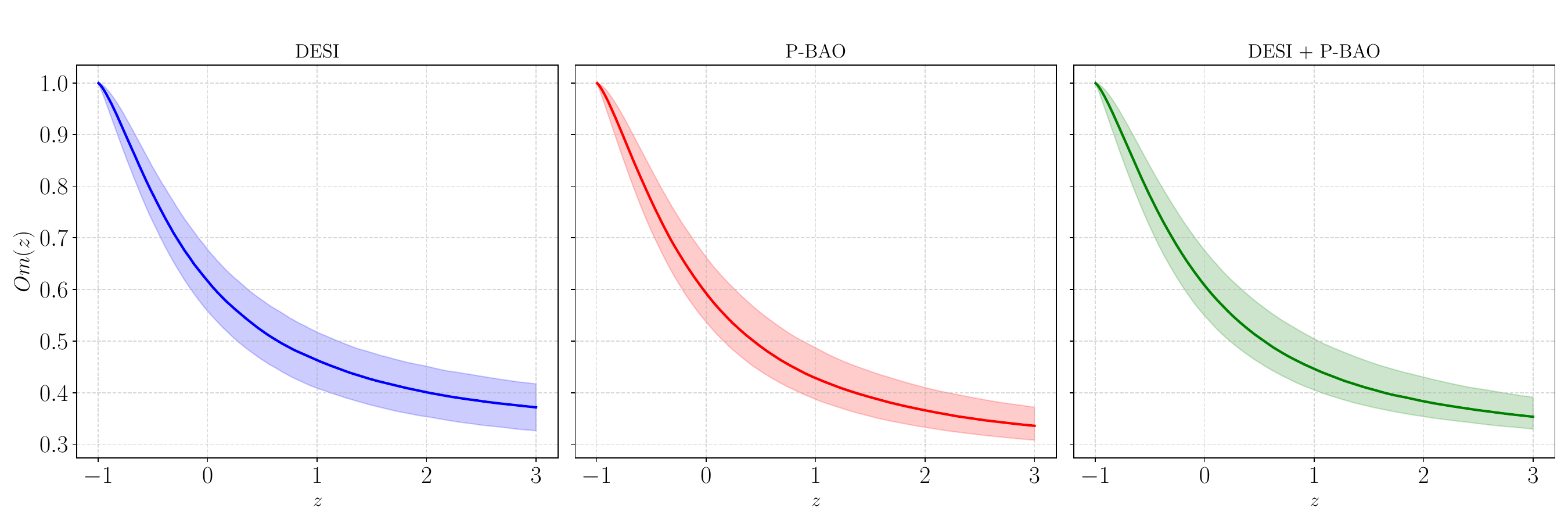}}
\caption{Evolution of $om$ diagnostic with redshift with best model parameters for the first model. The shaded regions here indicate the allowed region at 1$\sigma$ confidence level.}
\label{f23}
\end{figure*}
The $Om(z)$ diagnostic is a powerful geometrical tool used to distinguish dark energy models without requiring prior knowledge of the equation of state (EoS). In the standard $\Lambda$CDM model, $Om(z)$ is expected to remain constant across redshifts. Departures from this constant behaviour can hint at alternative dark energy dynamics:
\begin{itemize}
    \item \textbf{Constant} $Om(z)$: $\Lambda$CDM-like behavior.
    \item \textbf{Negative slope}: Quintessence-like behavior ($\omega > -1$).
    \item \textbf{Positive slope}: Phantom-like behavior ($\omega < -1$).
\end{itemize}
Fig. (\ref{f23}) clearly shows a negative slope for $Om(z)$ across all datasets — DESI, P-BAO, and their combination — especially in the low-redshift regime. This indicates that the reconstructed $f(Q)$ gravity model favours a quintessence-like dark energy behavior, where the effective EoS parameter satisfies $\omega_{\rm eff} > -1$. This observation is consistent with the results from the deceleration parameter $q(z)$ and effective EoS $\omega_{\rm eff}(z)$ analyses, which also suggest that the Universe is currently undergoing a phase of accelerated expansion driven by quintessence-type dynamics rather than a pure cosmological constant. Conclusively, the negative trend in $Om(z)$ provides further confirmation that the reconstructed $f(Q)$ model behaves differently from the standard $\Lambda$CDM and supports the viability of alternative gravity-based explanations for dark energy.
\subsection{Energy Conditions}
To further understand the physical viability of the reconstructed $f(Q)$ model, we analyze the classical energy conditions as functions of redshift $z$. These conditions, based on the effective energy density $\rho_{\text{eff}}$ and pressure $p_{\text{eff}}$, are given as:
\begin{itemize}
    \item \textbf{Weak Energy Condition (WEC)}: $\rho_{\text{eff}} \geq 0$, $\rho_{\text{eff}} + p_{\text{eff}} \geq 0 $ 
    \item \textbf{Null Energy Condition (NEC)}: $\rho_{\text{eff}} + p_{\text{eff}} \geq 0$
    \item \textbf{Strong Energy Condition (SEC)}: $\rho_{\text{eff}} + 3p_{\text{eff}} \geq 0$
    \item \textbf{Dominant Energy Condition (DEC)}: $\rho_{\text{eff}} - p_{\text{eff}} \geq 0 $
\end{itemize}
Here, $\rho_{\text{eff}}$ and $p_{\text{eff}}$ are defined through the modified Friedmann equations:
\begin{equation}
3H^2 = \rho_{\text{eff}},
\label{q1}
\end{equation}
\begin{equation}
2\dot{H} + 3H^2 = -p_{\text{eff}}.
\label{q2}
\end{equation}
\begin{figure*}[htb]
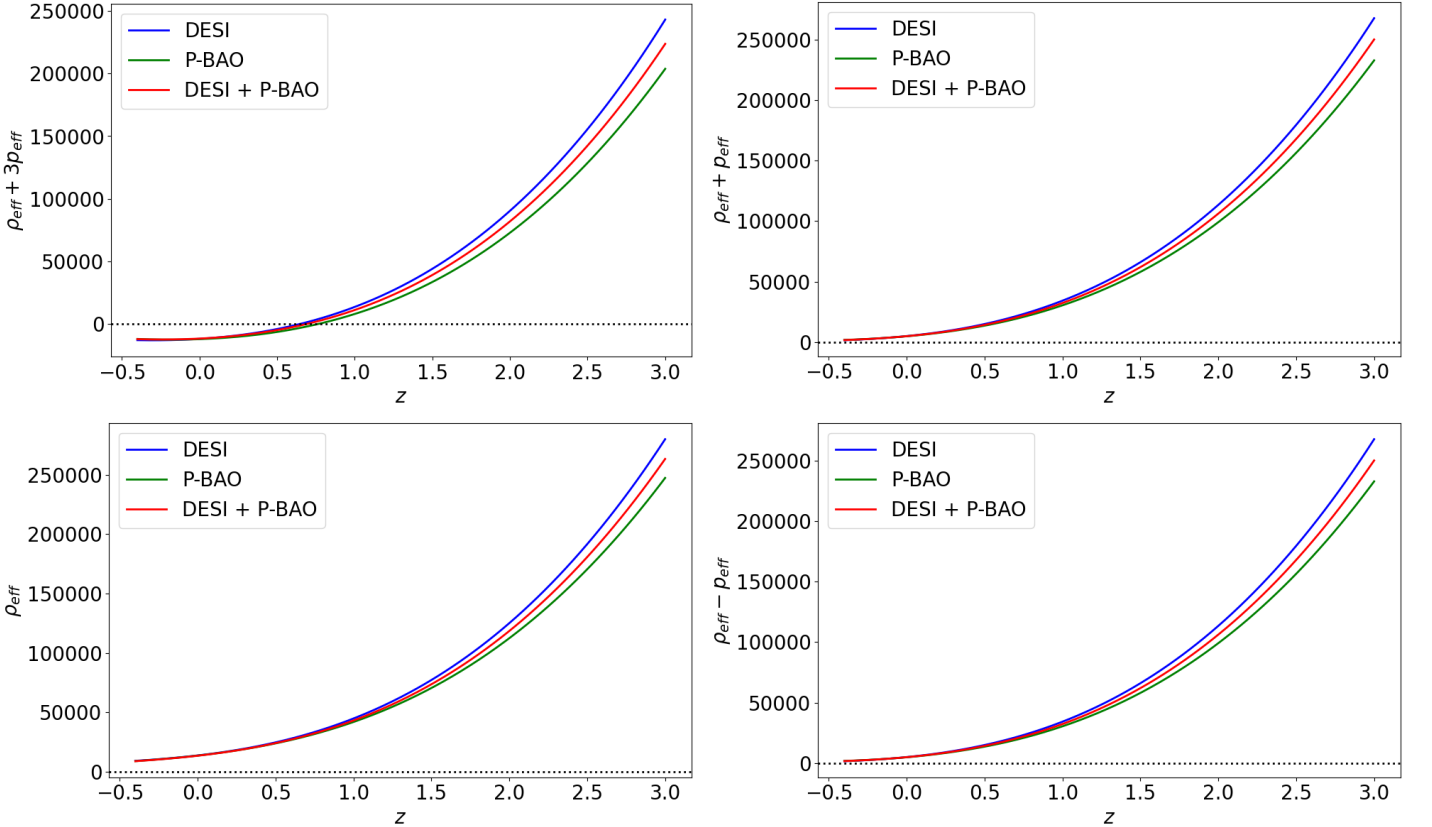

\centerline{
\includegraphics[width=.52\textwidth]{SEC}
\includegraphics[width=.52\textwidth]{NEC}}
\centerline{
\includegraphics[width=.52\textwidth]{DEC}
\includegraphics[width=.52\textwidth]{WEC}}
\caption{Evolution of deceleration parameter , effective EoS and $om$ diagnostic with redshift with best model parameters for the first model. The shaded regions here indicate the allowed region at 1$\sigma$ confidence level.}
\label{fe}
\end{figure*}
Based on our best-fit parameters from MCMC analysis using DESI and P-BAO datasets, the evolution of these energy conditions over redshift $z$  are evaluated by using Eqs. (\ref{e34}) and (\ref{e33}) with Eqs. (\ref{q1}) and (\ref{q2}), results for which are illustrated in Fig. (\ref{fe}). Results show that the Weak Energy Conditions  and Dominant Energy Conditions remain satisfied throughout the cosmic evolution, which is ensuring the positivity of effective energy density and confirming that the effective pressure never dominates over energy density. Conclusively this indicates a physically viable and non-exotic energy content in the Universe. The Null Energy Condition, often violated in phantom cosmologies, is preserved in our reconstructed $f(Q)$ model, supporting the interpretation of a smooth and stable dark energy sector with the absence of phantom-like behavior. Notably, the Strong Energy Condition is violated at low redshifts ($z \lesssim 1$), which aligns with the greatly observed late-time accelerated expansion of the Universe. However, at higher redshifts ($z \gg 1$) the SEC is restored, reflecting the transition from a decelerated matter-dominated phase to a late-time accelerated phase. These results demonstrate that the model successfully interpolates between a decelerated matter-dominated phase and a late-time accelerating phase of the Universe without invoking phantom energy or a cosmological constant, highlighting its consistency with observational features and physical viability.
\section{Conclusion}\label{s6}
In this work, we have obtained and analyzed a reconstructed form of the $f(Q)$ gravity model motivated by the New Agegraphic Dark Energy (NADE) scenario. This approach leverages the quantum gravitational origin of NADE to introduce a geometrically motivated modification to gravity, resulting in a model that naturally drives or explains the late-time accelerated expansion of the Universe without the introduction of a cosmological constant. Starting from a power-law scale factor of the form $a(t) = a_0 t^h$, we successfully derived an analytic form for the modified gravity function $f(Q)$ by employing the conformal time dependence inherent to NADE. The reconstructed function takes the form $f(Q) = Q + F(Q)$, with $F(Q) = \sqrt{Q} \left(\frac{a_0^2 6^h (h-1)^2 n^2 \left(\frac{h}{\sqrt{Q}}\right)^{2 h-1}}{h (2 h-1)}+C_1\right)$ carrying the dark energy modification in terms of the non-metricity scalar $Q$. This form is seen to naturally reduces to the symmetric teleparallel equivalent of General Relativity in the appropriate limit.\\
We constrained the free parameters of the model using observational data, including the recent only DESI BAO measurements, previous BAO compilations (P-BAO), and their combination for simplicity. A comprehensive Markov Chain Monte Carlo (MCMC) analysis was performed to extract the best-fit values and 1$\sigma$ confidence intervals for the model parameters $(H_0, \Omega_{m0}, a_0, h, n)$. The reconstructed Hubble parameter $H(z)$ from the best-fit model was found to be in excellent agreement with the observational $H(z)$ data, with high $R^2$ scores ($>0.98$) and low $\chi^2_{\min}$, AIC, and BIC values—demonstrating the statistical competitiveness of our model compared to the $\Lambda$CDM cosmology. To understand the underlying cosmological behaviour, we investigated several key diagnostics: the deceleration parameter $q(z)$, the effective equation of state $\omega_{\rm eff}(z)$, and the $Om(z)$ diagnostic. Our analysis shows a clear transition from a decelerated ($q > 0$) to an accelerated ($q < 0$) phase, with the present value of $q(0)$ lying within the range $[-0.5879, -0.3333]$, consistent with current observations. Similarly, $\omega_{\rm eff}(z)$ remains within the quintessence regime ($-1 < \omega_{\rm eff} < -1/3$), avoiding the phantom divide and supporting a stable and smooth late-time acceleration. The transition redshift $z_{\rm tr} \sim 0.5209$–$0.8126$ indicates the epoch of onset of cosmic acceleration. The $Om(z)$ diagnostic displays a negative slope in all cases, further confirming a quintessence-like evolution and distinguishing our model from $\Lambda$CDM, which would yield a constant $Om(z)$. We have also examined the energy conditions to ensure the physical viability of the reconstructed geometry. It was found that the Weak and Dominant Energy Conditions (WEC and DEC) remain satisfied throughout the cosmic evolution. The Null Energy Condition (NEC), which is often violated in phantom models, is seen to remain not violated here, strengthening the case for a stable dark energy component which is not phantom in nature. Notably, the Strong Energy Condition (SEC) is violated at low redshifts ($z \lesssim 1$), in alignment with the accelerating expansion of the universe, but is restored at higher redshifts—demonstrating a successful interpolation between matter-dominated and dark energy-dominated regimes.\\
The overall results suggest that our $f(Q)$ model, reconstructed via a physically motivated NADE approach, offers a compelling alternative to $\Lambda$CDM. It provides a consistent and observationally supported description of late-time cosmic acceleration without invoking a cosmological constant or phantom fields. Additionally, the model satisfies theoretical and observational requirements, including stability, physical energy conditions, and fit quality to $H(z)$ data. This framework can be further extended to study perturbations, structure formation, and inflationary dynamics, opening promising avenues for future investigation within the symmetric teleparallel gravity paradigm.


\begin{thebibliography}{200}

\bibitem{Riess1998}
A.~G.~Riess et al., ``Observational evidence from supernovae for an accelerating universe and a cosmological constant,'' \textit{Astron. J.} \textbf{116}, 1009 (1998), \href{https://arxiv.org/abs/astro-ph/9805201}{arXiv:astro-ph/9805201}.

\bibitem{Perlmutter1999}
S.~Perlmutter et al., ``Measurements of Omega and Lambda from 42 high-redshift supernovae,'' \textit{Astrophys. J.} \textbf{517}, 565 (1999), \href{https://arxiv.org/abs/astro-ph/9812133}{arXiv:astro-ph/9812133}.

\bibitem{Astier2006}
P.~Astier et al., ``The Supernova Legacy Survey: Measurement of $\Omega_M$, $\Omega_Lambda$ and $\omega$ from the first year data set,'' \textit{Astron. Astrophys.} \textbf{447}, 31 (2006), \href{https://arxiv.org/abs/astro-ph/0510447}{arXiv:astro-ph/0510447}.

\bibitem{Spergel2003}
D.~N.~Spergel et al., ``First-Year Wilkinson Microwave Anisotropy Probe (WMAP) Observations: Determination of Cosmological Parameters,'' \textit{Astrophys. J. Suppl. Ser.} \textbf{148}, 175 (2003), \href{https://arxiv.org/abs/astro-ph/0302209}{arXiv:astro-ph/0302209}.

\bibitem{Tegmark2004}
M.~Tegmark et al., ``Cosmological parameters from SDSS and WMAP,'' \textit{Phys. Rev. D} \textbf{69}, 103501 (2004), \href{https://arxiv.org/abs/astro-ph/0310723}{arXiv:astro-ph/0310723}.

\bibitem{Cole2005}
S.~Cole et al., ``The 2dF Galaxy Redshift Survey: power-spectrum analysis of the final dataset and cosmological implications,'' \textit{Mon. Not. R. Astron. Soc.} \textbf{362}, 505 (2005), \href{https://arxiv.org/abs/astro-ph/0501174}{arXiv:astro-ph/0501174}.

\bibitem{Eisenstein2005}
D.~J.~Eisenstein et al., ``Detection of the Baryon Acoustic Peak in the Large-Scale Correlation Function of SDSS Luminous Red Galaxies,'' \textit{Astrophys. J.} \textbf{633}, 560 (2005), \href{https://arxiv.org/abs/astro-ph/0501171}{arXiv:astro-ph/0501171}.

\bibitem{Weinberg1989}
S.~Weinberg, ``The cosmological constant problem,'' \textit{Rev. Mod. Phys.} \textbf{61}, 1 (1989).

\bibitem{Martin2012}
J.~Martin, ``Everything You Always Wanted To Know About The Cosmological Constant Problem (But Were Afraid To Ask),'' \textit{Comptes Rendus Physique} \textbf{13}, 566 (2012), \href{https://arxiv.org/abs/1205.3365}{arXiv:1205.3365 [astro-ph.CO]}.

\bibitem{Cai2007}
R.~G.~Cai, ``A Dark Energy Model Characterized by the Age of the Universe,'' \textit{Phys. Lett. B} \textbf{657}, 228 (2007), \href{https://arxiv.org/abs/0707.4049}{arXiv:0707.4049 [hep-th]}.

\bibitem{Karolyhazy1966}
F.~Karolyhazy, ``Gravitation and quantum mechanics of macroscopic objects,'' \textit{Il Nuovo Cimento A} \textbf{42}, 390 (1966).

\bibitem{Maziashvili2007a}
M.~Maziashvili, ``Space-time in light of Karolyhazy uncertainty relation,'' \textit{Int. J. Mod. Phys. D} \textbf{16}, 1531 (2007), \href{https://arxiv.org/abs/gr-qc/0612110}{arXiv:gr-qc/0612110}.

\bibitem{Maziashvili2007b}
M.~Maziashvili, ``Quantum fluctuations of spacetime and the dark energy problem,'' \textit{Phys. Lett. B} \textbf{652}, 165 (2007), \href{https://arxiv.org/abs/0705.0924}{arXiv:0705.0924 [gr-qc]}.

\bibitem{Wei2007}
H.~Wei and R.~G.~Cai, ``A New Model of Agegraphic Dark Energy,'' \textit{Phys. Lett. B} \textbf{660}, 113 (2008), \href{https://arxiv.org/abs/0708.0884}{arXiv:0708.0884 [astro-ph]}.

\bibitem{Wei2008}
H.~Wei and R.~G.~Cai, ``Interacting Agegraphic Dark Energy,'' \textit{Eur. Phys. J. C} \textbf{59}, 99 (2009), \href{https://arxiv.org/abs/0707.4052}{arXiv:0707.4052 [hep-th]}.

\bibitem{Sheykhi2009}
A.~Sheykhi, ``Interacting New Agegraphic Dark Energy in Brans-Dicke Theory,'' \textit{Phys. Lett. B} \textbf{681}, 205 (2009), \href{https://arxiv.org/abs/0907.5458}{arXiv:0907.5458 [hep-th]}.

\bibitem{Karami2010}
K.~Karami and A.~Abdolmaleki, ``New agegraphic dark energy in Brans-Dicke theory,'' \textit{Phys. Scr.}, \textbf{81}, 055901 (2010).

\bibitem{Karami2011a}
K.~Karami and M.~S.~Khaledian, ``Reconstructing $f(R)$ modified gravity from entropy-corrected versions of holographic and new agegraphic dark energy models,'' \textit{JHEP}, \textbf{03}, 086 (2011).

\bibitem{Karami2011b}
K.~Karami et al., ``Reconstructing modified gravity models from the entropy-corrected agegraphic dark energy,'' \textit{Phys. Lett. B}, \textbf{686}, 216 (2010).


\bibitem{Capozziello2006}
S.~Capozziello, V.~F.~Cardone, and A.~Troisi, ``Dark energy and dark matter as curvature effects,'' \textit{JCAP} \textbf{0608}, 001 (2006), \href{https://arxiv.org/abs/astro-ph/0602349}{arXiv:astro-ph/0602349}.

\bibitem{Nojiri2006recon}
S.~Nojiri and S.~D.~Odintsov, ``Modified f(R) gravity consistent with realistic cosmology: From matter dominated epoch to dark energy universe,'' \textit{Phys. Rev. D} \textbf{74}, 086005 (2006), \href{https://arxiv.org/abs/hep-th/0608008}{arXiv:hep-th/0608008}.

\bibitem{Sotiriou2010}
T.~P.~Sotiriou and V.~Faraoni, ``$f(R)$ theories of gravity,'' \textit{Rev. Mod. Phys.} \textbf{82}, 451 (2010), \href{https://arxiv.org/abs/0805.1726}{arXiv:0805.1726 [gr-qc]}.

\bibitem{Jimenez2018}
J.~B.~Jiménez, L.~Heisenberg, and T.~Koivisto, ``Coincident General Relativity,'' \textit{Phys. Rev. D} \textbf{98}, 044048 (2018), \href{https://arxiv.org/abs/1710.03116}{arXiv:1710.03116 [gr-qc]}.

\bibitem{Mandal2020}
S.~Mandal, P.~K.~Sahoo, and J.~R.~L.~Santos, ``Cosmography in f(Q) gravity,'' \textit{Phys. Rev. D} \textbf{102}, 024057 (2020), \href{https://arxiv.org/abs/2001.02357}{arXiv:2001.02357 [gr-qc]}.

\bibitem{Dunsby2010}
P.~K.~S.~Dunsby, E.~Elizalde, R.~Goswami, S.~Odintsov, and D.~S.~Gomez, ``On the LCDM Universe in f(R) gravity,'' \textit{Phys. Rev. D} \textbf{82}, 023519 (2010), \href{https://arxiv.org/abs/1005.2205}{arXiv:1005.2205 [gr-qc]}.

\bibitem{Goheer2009}
N.~Goheer, R.~Goswami, and P.~K.~S.~Dunsby, ``Dynamics of $f(G)$ cosmology,'' \textit{Phys. Rev. D} \textbf{79}, 121301 (2009), \href{https://arxiv.org/abs/0904.2559}{arXiv:0904.2559 [gr-qc]}.

\bibitem{Baffou2017}
E.~H.~Baffou, M.~J.~S.~Houndjo, and J.~Tossa, ``Reconstruction of $f(\tau,T)$ gravity in different cosmological models,'' \textit{Eur. Phys. J. C} \textbf{77}, 708 (2017), \href{https://arxiv.org/abs/1706.08842}{arXiv:1706.08842 [gr-qc]}.

\bibitem{Saha2025}
P.~Saha and P.~Rudra, ``A Cosmological Holographic Reconstruction of $f(Q)$ Theory,'' \textit{arXiv:2407.01870v2} (2025), \href{https://arxiv.org/abs/2407.01870}{arXiv:2407.01870 [gr-qc]}.

\bibitem{f1}
 S. Nojiri and S.D. Odintsov, Introduction to modified gravity and gravitational alternative
 for dark energy, Int. J. Geom. Meth. Mod. Phys. 4 (2007) 115
 
 
\bibitem{refS66} Burnham, K. P., \& Anderson, D. R. (2004). Multimodel inference: Understanding AIC and BIC in model selection. \textit{Sociological Methods \& Research}, \textit{33}(2), 261--304. \href{https://doi.org/10.1177/0049124104268644}{https://doi.org/10.1177/0049124104268644}

\bibitem{refS68} Liddle, A. R. (2004). How many cosmological parameters? \textit{Monthly Notices of the Royal Astronomical Society}, \textit{351}(3), L49--L53. \href{https://arxiv.org/abs/astro-ph/0401198}{astro-ph/0401198}

\bibitem{refS64} Schwarz, G. (1978). Estimating the dimension of a model. \textit{The Annals of Statistics}, \textit{6}(2), 461--464. \href{https://doi.org/10.1214/aos/1176344136}{https://doi.org/10.1214/aos/1176344136}

\bibitem{ref98} Adame, A. G., et al. (2024). DESI 2024 VI: Cosmological Constraints from the Measurements of Baryon Acoustic Oscillations. \textit{arXiv}. \href{https://arxiv.org/abs/2404.03002}{https://arxiv.org/abs/2404.03002}

\bibitem{d113} Gaztanaga, E., Cabre, A., \& Hui, L. (2009). Clustering of Luminous Red Galaxies IV: Baryon Acoustic Peak in the Line-of-Sight Direction and a Direct Measurement of H(z). \textit{Monthly Notices of the Royal Astronomical Society}, \textit{399}, 1663–1680. \href{https://arxiv.org/abs/0807.3551}{arXiv:0807.3551}

\bibitem{d114} Oka, A., Saito, S., Nishimichi, T., Taruya, A., \& Yamamoto, K. (2014). Simultaneous constraints on the growth of structure and cosmic expansion from the multipole power spectra of the SDSS DR7 LRG sample. \textit{Monthly Notices of the Royal Astronomical Society}, \textit{439}, 2515–2530. \href{https://arxiv.org/abs/1310.2820}{arXiv:1310.2820}

\bibitem{d115} Wang, Y., et al. (2017). The clustering of galaxies in the completed SDSS-III Baryon Oscillation Spectroscopic Survey: tomographic BAO analysis of DR12 combined sample in configuration space. \textit{Monthly Notices of the Royal Astronomical Society}, \textit{469}(3), 3762–3774. \href{https://arxiv.org/abs/1607.03154}{arXiv:1607.03154}

\bibitem{d116} Chuang, C.-H., \& Wang, Y. (2013). Modeling the Anisotropic Two-Point Galaxy Correlation Function on Small Scales and Improved Measurements of H(z), DA(z), and $\beta$(z) from the Sloan Digital Sky Survey DR7 Luminous Red Galaxies. \textit{Monthly Notices of the Royal Astronomical Society}, \textit{435}, 255–262. \href{https://arxiv.org/abs/1209.0210}{arXiv:1209.0210}

\bibitem{d117} Anderson, L., et al. (2014). The clustering of galaxies in the SDSS-III Baryon Oscillation Spectroscopic Survey: baryon acoustic oscillations in the Data Releases 10 and 11 Galaxy samples. \textit{Monthly Notices of the Royal Astronomical Society}, \textit{441}(1), 24–62. \href{https://arxiv.org/abs/1312.4877}{arXiv:1312.4877}

\bibitem{d118} Zhao, G.-B., et al. (2019). The clustering of the SDSS-IV extended Baryon Oscillation Spectroscopic Survey DR14 quasar sample: a tomographic measurement of cosmic structure growth and expansion rate based on optimal redshift weights. \textit{Monthly Notices of the Royal Astronomical Society}, \textit{482}(3), 3497–3513. \href{https://arxiv.org/abs/1801.03043}{arXiv:1801.03043}

\bibitem{d119} Busca, N. G., et al. (2013). Baryon Acoustic Oscillations in the Ly-$\alpha$ forest of BOSS quasars. \textit{Astronomy \& Astrophysics}, \textit{552}, A96. \href{https://arxiv.org/abs/1211.2616}{arXiv:1211.2616}

\bibitem{d120} Font-Ribera, A., et al. (2014). Quasar-Lyman $\alpha$ Forest Cross-Correlation from BOSS DR11: Baryon Acoustic Oscillations. \textit{Journal of Cosmology and Astroparticle Physics}, \textit{2014}(05), 027. \href{https://arxiv.org/abs/1311.1767}{arXiv:1311.1767}

\bibitem{d121} du Mas des Bourboux, H., et al. (2017). Baryon acoustic oscillations from the complete SDSS-III Ly$\alpha$-quasar cross-correlation function at z = 2.4. \textit{Astronomy \& Astrophysics}, \textit{608}, A130. \href{https://arxiv.org/abs/1708.02225}{arXiv:1708.02225}

\bibitem{d5} Alam, S., et al. (2017). The clustering of galaxies in the completed SDSS-III Baryon Oscillation Spectroscopic Survey: cosmological analysis of the DR12 galaxy sample. \textit{Monthly Notices of the Royal Astronomical Society}, \textit{470}(3), 2617–2652. \href{https://arxiv.org/abs/1607.03155}{arXiv:1607.03155}

\bibitem{d10} Chuang, C.-H., et al. (2013). The clustering of galaxies in the SDSS-III Baryon Oscillation Spectroscopic Survey: single-probe measurements and the strong power of normalized growth rate on constraining dark energy. \textit{Monthly Notices of the Royal Astronomical Society}, \textit{433}, 3559. \href{https://arxiv.org/abs/1303.4486}{arXiv:1303.4486}

\bibitem{d74} Blake, C., et al. (2012). The WiggleZ Dark Energy Survey: Joint measurements of the expansion and growth history at $z < 1$. \textit{Monthly Notices of the Royal Astronomical Society}, \textit{425}, 405–414. \href{https://arxiv.org/abs/1204.3674}{arXiv:1204.3674}

\bibitem{d79} Neveux, R., et al. (2020). The completed SDSS-IV extended Baryon Oscillation Spectroscopic Survey: BAO and RSD measurements from the anisotropic power spectrum of the quasar sample between redshift 0.8 and 2.2. \textit{Monthly Notices of the Royal Astronomical Society}, \textit{499}(1), 210–229. \href{https://arxiv.org/abs/2007.08999}{arXiv:2007.08999}


\bibitem{planck2018} Planck Collaboration, Aghanim, N., Akrami, Y., et al. (2020). Planck 2018 results. VI. Cosmological parameters. \textit{Astronomy \& Astrophysics}, \textit{641}, A6. \href{https://arxiv.org/abs/1807.06209}{arXiv:1807.06209}

\bibitem{verde2019tensions} Verde, L., Treu, T., \& Riess, A. G. (2019). Tensions between the early and late Universe. \textit{Nature Astronomy}, \textit{3}, 891–895. \href{https://arxiv.org/abs/1907.10625}{arXiv:1907.10625}

\bibitem{farooq2013} Farooq, O., \& Ratra, B. (2013). Hubble parameter measurement constraints on the cosmological deceleration–acceleration transition redshift. \textit{The Astrophysical Journal Letters}, \textit{766}(1), L7. \href{https://arxiv.org/abs/1301.5243}{arXiv:1301.5243}

\bibitem{xu2012} Zheng, W., Li, H., Xia, J.-Q., Wan, Y.-P., Li, S.-Y., \& Li, M. (2014). Constraints on cosmological models from Hubble parameters measurements. \textit{International Journal of Modern Physics D}, \textbf{23}(05), 1450051. \href{https://doi.org/10.1142/s0218271814500515}{https://doi.org/10.1142/s0218271814500515}.
 







\end{thebibliography}
\end{document}